\documentclass[reprint, superscriptaddress, amsmath, amssymb, aps, floatfix, 
prx]{revtex4-2}
\usepackage[latin9]{inputenc}
\usepackage{verbatim}
\usepackage{amsmath}
\usepackage{amssymb}
\usepackage{graphicx}

\usepackage{amsthm}
\usepackage{fancyhdr}
\usepackage{epsfig}
\usepackage{bbm}
\usepackage{subfigure}
\usepackage{float}
\usepackage{xcolor}
\usepackage{tikz}
\usepackage[normalem]{ulem}
\usepackage[colorlinks=true,linkcolor=blue,citecolor=blue]{hyperref}
\usepackage{multirow}

\newcommand{\ket}[1]{|{#1})}
\newcommand{\bra}[1]{({#1}|}
\newcommand{\kets}[1]{|{#1}\rangle}
\newcommand{\bras}[1]{\langle{#1}|}

\begin{document}

\title{The chiral random walk: A quantum-inspired framework for odd diffusion}

\author{Jan W{\'o}jcik}
\email{jan.wojcik@phdstud.ug.edu.pl}
\affiliation{Institute of Theoretical Physics and Astrophysics, University 
of Gda{\'n}sk, 80-308 Gda{\'n}sk, Poland}
\affiliation{Institute of Spintronics and Quantum Information, Faculty of 
Physics and Astronomy, Adam Mickiewicz University, 61-614 Pozna{\'n}, Poland}
\author{Erik Kalz}
\email{erik.kalz@uni-potsdam.de}
\affiliation{Institute of Physics and Astronomy, University of Potsdam,
14476 Potsdam, Germany}

\begin{abstract}
Chirality in active and passive fluids gives rise to \textit{odd} transport 
properties, most notably the emergence of robust edge currents that defy 
standard dissipative dynamics. While these phenomena are well-described by 
continuum hydrodynamics, a microscopic framework connecting them to their 
topological origins has remained elusive. Here, we present a lattice model 
for an isotropic chiral random walk that bridges the gap between classical 
stochastic diffusion and unitary quantum evolution. By equipping the walker 
with an internal degree of freedom and a tunable chirality parameter, $p$, 
we interpolate between a standard diffusive random walk and a deterministic, 
topologically non-trivial quantum walk. We show that the topological protection 
characteristic of the unitary limit ($p=1$) remarkably persists into the 
dissipative regime ($p<1$). This correspondence allows us to theoretically 
ground the robustness of edge flows in classical chiral systems using the 
bulk-boundary correspondence of Floquet topological insulators. Our results 
provide a discrete microscopic description for odd diffusion, offering a 
powerful toolkit to predict transport in confined geometries and disordered 
chiral media.
\end{abstract}

\maketitle

\textit{Introduction.} The problem of a discrete random walk was introduced 
by the famous \textit{drunkard's walk} in an exchange of letters between 
Pearson and Strutt, better known as Lord Rayleigh, in three consecutive volumes 
of \textit{Nature} in 1905 \cite{pearson1905problem_1, rayleigh1905problem, 
pearson1905problem_2}. In the third letter, Pearson concluded that ``\textit{the 
most probable place to find a drunken man (...) is somewhere near his 
starting point}'' \cite{pearson1905problem_2}. A full, formal solution of the 
random walk was already provided two months after the original problem 
statement by Kluyver \cite{kluyver1905local}. The random walk problem was 
independently used in the early theory of diffusion; both Einstein 
\cite{einstein1905uber} and Smoluchowski \cite{smoluchowski1906zur} essentially 
argued with it in their seminal works, Smoluchowski later also acknowledged 
the connection to the Pearson problem \cite{smoluchowksi1916drei}. Since then, 
the random walk has proven to be of great historical relevance, guiding 
developments in polymer theory, percolation theory, and many other fields 
\cite{chandrasekhar1943stochastic, hughes1995random1}. The idea of the random 
walk was generalized to arbitrary dimensions \cite{polya1921aufgabe} or, more 
recently, led to anomalous diffusion when considering arbitrary jump size and 
length distributions \cite{metzler2000random}. In 1993, Aharonov et \textit{al.} 
\cite{aharonov1993quantum} broke the classical paradigm within the random walk 
model by introducing the \textit{quantum random walk}, which harnesses quantum 
interference through unitary evolution, utilizing an internal degree of freedom 
(IDF) to facilitate the superposition, for a review see \cite{Kempe2003quantum}.

\begin{figure*}[t]
\includegraphics[width=\linewidth]{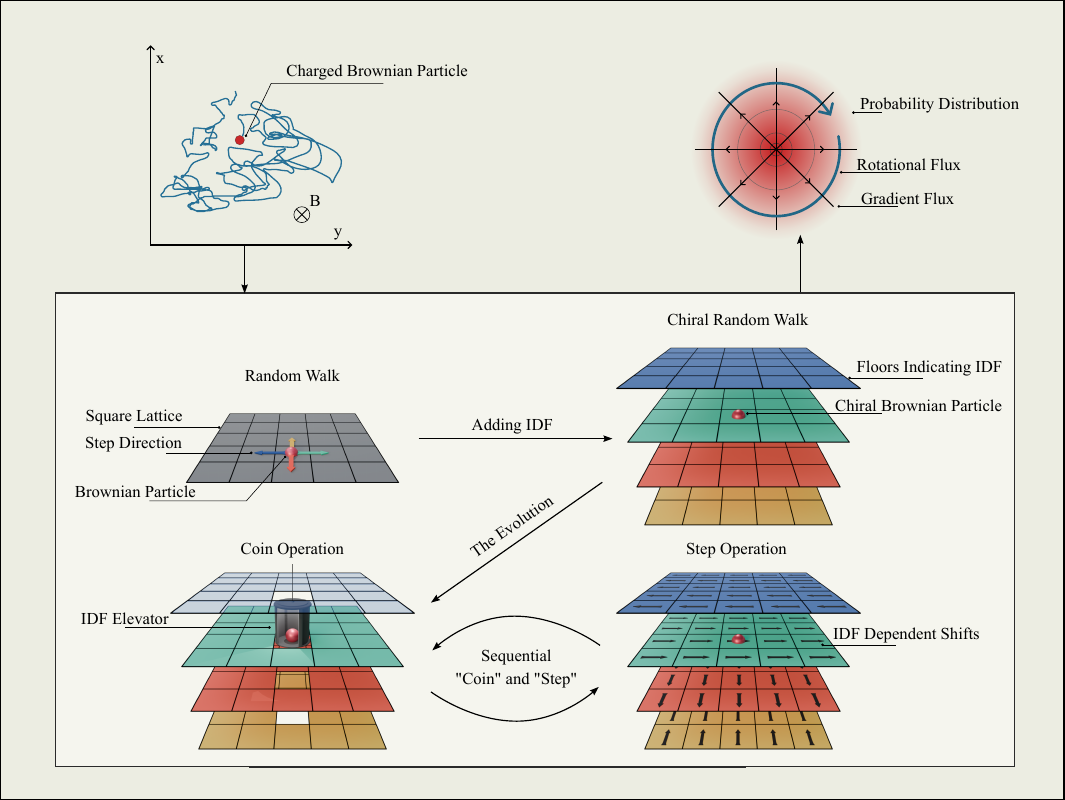}
\caption{\textbf{Conceptual scheme of the chiral random walk model}. 
A charged Brownian particle in a magnetic field experiences Lorentz forces 
that induce chiral motion. Its diffusive motion, however, cannot be captured 
by a standard random walk model on a lattice. To model such systems, we 
introduce an internal degree of freedom (IDF), represented schematically as 
floors in a building. The time evolution then decomposes into two sequential 
operations: the coin operation, which acts on the IDF, similar to a lift 
between the floors, while the step operation shifts the particle's position on 
the lattice according to which floor, i.e., which IDF state ($\rightarrow, 
\leftarrow, \uparrow, \downarrow$), it currently has. This structure enables 
the model to incorporate passive chirality, resulting in a probability 
distribution that exhibits not only standard gradient flux but also rotational 
flux, the hallmark of odd diffusion that cannot be reproduced by classical 
random walks.}
\label{fig:crw}
\end{figure*}

Much to our surprise, however, a rather fundamental property of physical 
systems remains almost overlooked in the literature---chiral random walks. 
Chirality in motion typically originates from a broken parity or time-reversal 
symmetry of the underlying process and predominantly appears for living 
microorganisms. Already in 1901, Jennings noted ``\textit{On the significance 
of the spiral swimming of organisms}'' \cite{jennings1901significance}, a topic 
that in recent years also has attracted the interest of physicists 
\cite{lowen2016chirality, liebchen2022chiral}. Investigations range from 
artificial colloidal systems \cite{kummel2013circular, wykes2016dynamic}, 
over self-propelled chiral agents---bacteria and algae \cite{frymier1995three, 
lauga2006swimming, petroff2015fast, drescher2009dancing} to quasi-continuous 
chiral fluids \cite{li2024robust}. Chirality, however, does not only emerge 
within active systems. Further, external fields can break the time-reversal 
symmetry and thus give rise to typical chiral behaviour of the tracer particles. 
The paradigmatic example is a magnetic field to which dipolar agents react in 
the perpendicular plane, and when the field is rotated, they form chiral 
colloidal fluids \cite{soni2019odd, massana2021arrested}. If the colloidal 
particles instead carry a charge, the magnetic field induces a Lorentz force, 
which eventually also results in chiral trajectories of the agents 
\cite{lemons1999brownian, czopnik2001brownian, jimenez2006brownian}. Notably, 
Lorentz forces in colloidal matter have recently found a renewed interest in 
the coarse-grained description of skyrmionic spin vortex structures that 
follow the same time-evolution in the overdamped limit \cite{troncoso2014brownian, 
reichhardt2022statics}. 

In the most recent years, these seemingly different systems have found a 
unifying terminology and started to be investigated from a joint perspective. 
There, chiral motion is described in terms of transport tensors with 
antisymmetric off-diagonal elements. These elements behave odd under inverting 
the direction of chirality, which has become the namesake for such \textit{odd} 
systems. In two spatial dimensions, oddness does not conflict with isotropy 
and in fact, odd transport tensor represent the most general isotropic 
description of a system \cite{dresselhaus2025anomalous}. Chiral fluids, for 
example, are described by odd viscosity on the continuum level 
\cite{fruchart2023odd}, networks with transverse forces (a local source of 
chirality) are described by odd elasticity \cite{scheibner2020odd}. On the 
discrete level, chiral agents perform odd-diffusive motion \cite{hargus2021odd, 
kalz2022collisions, kalz2024oscillatory} and react with odd mobility to forces 
in the system \cite{poggioli2023odd, caprini2025bubble, kalz2025reversal}. 
Despite the efforts on the continuous description of chiral processes, little
work has been done on a discrete, chiral random walk (CRW) model. To our 
knowledge, only active chirality has been dealt with from a discrete perspective, 
with the intrinsic feature that chirality was represented by a drift preference 
and thus requires a nonequilibrium setting of an active walker 
\cite{hargus2021odd}. The difficulty in relaxing that property can be illustrated 
by considering the process of overdamping the Brownian particle with Lorentz 
force \cite{chun2018emergence, vuijk2019anomalous, park2021thermodynamic}, a 
prototypical passive chiral process. The broken time-reversal symmetry on the 
velocity process, as induced by the magnetic field, has to be accounted for in 
the limit of vanishing inertia---which is precisely what the antisymmetric 
odd-diffusion tensor does. However, the price is a Gaussian, but nonwhite noise, 
which inherits the broken time-reversal symmetry in its fluctuations 
\cite{chun2018emergence}. That, however, can hardly be understood from a 
discrete perspective: the seemingly intuitive way to include chirality via an 
orthogonal bias in the step fails for an arbitrary chirality 
\cite{hijikata2015markovian}. 

Chiral systems, active and passive, have a surprising property. While in the 
bulk, the dynamics of an odd-diffusive tracer appear divergence-free and 
virtually indistinguishable from ordinary diffusion, when boundaries are 
introduced, persistent edge flows emerge. These edge flows are not only 
remarkably robust against disorder \cite{souslov2019topological,caporusso2024phase}, 
but have started to become an experimental indicator of chirality 
\cite{soni2019odd, yang2020robust, yang2021topologically, li2024robust, 
nelson2025topological}. This robustness has prompted an analogy with topological 
phenomena in quantum systems, where the celebrated bulk-boundary correspondence 
guarantees protected edge transport \cite{hasan2010colloquium}. Consequently, 
efforts have been directed toward connecting odd systems to the quantum Hall 
effect \cite{souslov2017topological, souslov2019topological, lou2022odd}, 
thereby opening the door to applying the extensive toolkit developed in quantum 
mechanics to analyse topological behaviour in classical, dissipative settings.
A growing body of both theoretical and experimental work has strengthened this 
connection \cite{shankar2022topological, sone2020exceptional}, often by 
mapping steady-state solutions of dissipative systems to zero-energy eigenstates 
of quantum Hamiltonians with nontrivial topological invariants 
\cite{dasbiswas2018topological}. Establishing such mappings is far from trivial, 
as dissipative dynamics are inherently probabilistic, in contrast to the 
deterministic evolution of closed quantum systems. Nevertheless, once a reliable 
mapping is achieved, the predictive power of quantum theory can be harnessed
to explain, and even anticipate, complex features of odd systems, such as 
robust edge transport.

In this work, we introduce a discrete random walk model for a (passive) chiral 
tracer that captures the essential features of odd diffusion. To achieve this, 
we adopt the concept of an IDF, akin to that used in discrete-time quantum 
walks. This IDF acts as a directional guide for the particle's motion. 
Crucially, the evolution is governed by a single parameter $p$ ($0 \leq p \leq 1$) 
that interpolates between stochastic and deterministic dynamics. At one extreme, 
when $p=0$, we recover the classical random walk. For $0<p<1$, the walker shows 
the characteristics of a chiral tracer with odd diffusion. At the other 
extreme, for $p=1$, the dynamics becomes fully deterministic. As such, it is 
governed by a bistochastic, unitary evolution matrix and can be mapped to a 
quantum walk. The latter allows us to apply quantum mechanical tools, and 
specifically enables us to probe the emergence of topologically protected edge 
transport in chiral, diffusive systems.

\textit{Model.} We outline the model by starting to construct the random walk 
on a square lattice $a\mathbb{Z}^2$, where $a$ is the lattice constant, that 
we set to one for simplicity. The time evolution is discrete, with integer 
multiples of the unit timescale $\tau$, which we also set to one. At time-step 
$t$, the walker has a position $\mathbf{X}_t = (X_t,Y_t) \in  \mathbb{Z}^2$. 
We denote by $p_\mathbf{x}(t)$ the probability of the walker to be at position 
$\mathbf{X}_t$, at time $t$, i.e., $p_{\mathbf{x}}(t) =  
\mathrm{Prob}\{\mathbf{X}_t = \mathbf{x}\}$. The probability vector, akin to 
the probability amplitude vector in quantum mechanics, can be written as
\begin{equation} 
    \ket{\rho(t)} = \sum_{\mathbf{x}} p_\mathbf{x}(t) \ket{\mathbf{x}},
\end{equation}
where $\ket{\mathbf{x}} = \ket{x}\ket{y}$ are the spatial basis eigenvectors, 
e.g., $\ket{x=0} = (1, 0, \ldots, 0)^\mathrm{T}$ where $(\cdot)^\mathrm{T}$ 
denotes the usual matrix transposition. 

The discrete-time random walk evolves in finite steps according to
\begin{equation}
    \ket{\rho(t+1)} = U \ket{\rho(t)}
\end{equation}
where $U$ is the stochastic evolution operator. In the standard random walk, 
$U$ shifts the tracer to nearest-neighbour positions with equal probability
\begin{align}
    U \ket{\rho(t)} &= \sum_{\mathbf{x}} p_{\mathbf{x}}(t) U\ket{\mathbf{x}} 
    \nonumber \\
    &= \sum_{\mathbf{x}} \sum_{\boldsymbol{\sigma}} p_{\mathbf{x}}(t) \, 
    \frac{1}{4} \ket{\mathbf{x} + \boldsymbol{\sigma}},
    \label{eq_classical_random_walk}
\end{align}
where $\boldsymbol{\sigma} = (\sigma_x, \sigma_y)$ and $\sigma_i\in\{+1, -1\}$, 
$i=x,y$.

This time evolution resembles Markovian dynamics, as after each time step, 
the information about the previous position of the tracer is lost 
\cite{hughes1995random1}. To generalize this model to include chirality, 
this introduces a conceptual difficulty, as one step should introduce a bias 
for the next one. For example, in clockwise chiral motion, a step to the right 
should increase the probability of a subsequent downward step.

For the CRW, we have to extend the state of the particle with an IDF 
$\mathcal{D}\in\{\rightarrow, \leftarrow, \uparrow, \downarrow\}$. The full-time 
evolution of the process acts on the product-space $\mathbb{Z}^2 \otimes 
\mathcal{D}$. The probability of finding the particle at position $\mathbf{X}_t$ 
with internal degree $\mathcal{D}_t$ at time $t$ now is the joint probability 
$p_{\mathbf{x},d}(t) =  \mathrm{Prob}\{\mathbf{X}_t = \mathbf{x}, 
\mathcal{D}_t = d\}$. Note that marginalizing the internal degree gives the 
previous probability $p_\mathbf{x}(t)=\sum_{d} p_{\mathbf{x},d}(t)$. The 
probability vector now incorporates the internal degree via
\begin{equation}
\ket{\rho(t)} = \sum_{\mathbf{x}, d} p_{\mathbf{x}, d}(t)\, \ket{\mathbf{x}} 
\otimes \ket{d}.
\end{equation}
The four basis vectors of $\mathcal{D}$  are given by $\ket{\rightarrow} = 
(1,0,0,0)^\mathrm{T}$, $\ket{\leftarrow} = (0,1,0,0)^\mathrm{T}$, 
$\ket{\uparrow} = (0,0,1,0)^\mathrm{T}$, $\ket{\downarrow} = (0,0,0,1)^\mathrm{T}$. 
We also denote the product state as $\ket{\mathbf{x}, d}$. 

Inspired by the time-evolution in the quantum walk model \cite{Kempe2003quantum}, 
which also utilizes an IDF, we propose a decomposition of $U$ as 
\begin{equation}\label{evolution}
    U = S (I \otimes C),
\end{equation}
which is a product of a step operator $S$ that accounts for the topological 
structure of the lattice, and a coin operator $C$ that governs the transition 
probabilities between internal states. The identity $I$ here acts on the 
spatial dimensions. The step operator $S$ is a conditional permutation matrix 
that shifts the particle according to its internal state. For the 
nearest-neighbour walk on the square lattice that we consider here, the four 
possible steps are given by
\begin{subequations}
\label{step_operator}
\begin{align}
    \label{step_operator_1}
    S\ket{x,y,\rightarrow} &= \ket{x+1,y,\rightarrow}, \\
    S\ket{x,y,\leftarrow} &= \ket{x-1,y,\leftarrow}, \\
    S\ket{x,y,\uparrow} &= \ket{x,y+1,\uparrow}, \\
    \label{step_operator_4}
    S\ket{x,y,\downarrow} &= \ket{x,y-1,\downarrow}.
\end{align}
\end{subequations}
The coin operator $C$ determines the transition probabilities and is a 
stochastic $4 \times 4$ matrix in our case, given by 
\begin{equation}\label{coin_rw}
    C_{\text{rand}} = \frac{1}{4}\begin{pmatrix}
        1 && 1 && 1 && 1 \\
        1 && 1 && 1 && 1 \\
        1 && 1 && 1 && 1 \\
        1 && 1 && 1 && 1 
    \end{pmatrix}.
\end{equation}
This choice resembles the time-evolution of Eq.~\eqref{eq_classical_random_walk} 
as it randomizes the internal state after each step. In general, $C$ can also 
be position-dependent (such that in Eq.~\ref{evolution} $ I\otimes C 
\rightarrow \sum_{\mathbf {x} }\ket{\mathbf{x}}\bra{\mathbf{x}}\otimes 
C_{\mathbf{x}}$), to account for spatial inhomogeneities or boundaries in the 
system. For example, we can model a reflective boundary condition as 
introduced by a hard wall by setting the $C_\mathbf{x}$ at the boundaries to 
\begin{equation}
\label{refelctive_coin}
    C_{\text{refl} } =\begin{pmatrix}
        0 && 1 && 0 && 0 \\
        1 && 0 && 0 && 0 \\
        0 && 0 && 0 && 1 \\
        0 && 0 && 1 && 0 
    \end{pmatrix}.
\end{equation}

We can now use this internal state of the walker to implement a stochastic 
chiral walk. To gain intuition, we start by demonstrating that a fully 
deterministic clockwise walk,
\begin{equation}
\label{running_in_circles}
    \begin{array}{ccc}
      \ket{x,y,\rightarrow} & \longrightarrow  & \ket{x+1,y,\downarrow}\\
      \big\uparrow &  & \big\downarrow \\
       \ket{x,y-1,\uparrow}\ & \longleftarrow & \ket{x+1,y-1,\leftarrow}\ 
    \end{array},
\end{equation}
can be obtained by a coin operator of the form 
\begin{equation}
\label{coin_chiral}
    C_{\text{chir}} =  \begin{pmatrix}
        0 && 0 && 1 && 0 \\
        0 && 0 && 0 && 1 \\
        0 && 1 && 0 && 0 \\
        1 && 0 && 0 && 0 
    \end{pmatrix}.
\end{equation}

We are now in the position to introduce the  \textit{chiral random walk}, which 
combines the classical random walk model, obtained by a fully mixing coin 
operator $C_\mathrm{rand}$ of Eq.~\eqref{coin_rw} with probability $1-p$, with 
the chiral coin $C_\mathrm{chir}$ of Eq.~\eqref{coin_chiral}, with probability 
$p$, where $0\leq p \leq 1$. The CRW coin thus takes the form 
\begin{equation}
\label{coin_chiral_random}
    C_{\text{CRW}} = \frac{1-p}{4}\begin{pmatrix}
        1 && 1 && 1 && 1 \\
        1 && 1 && 1 && 1 \\
        1 && 1 && 1 && 1 \\
        1 && 1 && 1 && 1 
    \end{pmatrix} + p \begin{pmatrix}
        0 && 0 && 1 && 0 \\
        0 && 0 && 0 && 1 \\
        0 && 1 && 0 && 0 \\
        1 && 0 && 0 && 0 
    \end{pmatrix}.
\end{equation}
As an example, we consider an initial state $\ket{\rho(t=0)} =  
\ket{\mathbf{x}=\mathbf{0}, \rightarrow}$ that evolves after one time step to 
the state $\ket{\rho(t=1)} =  \ket{x=1, y=0} \otimes (1-p,1-p, 1-p, 
1 + 3p)^\mathrm{T}/4$. Thus, after a step to the right, for $p>0$, the next step 
in the downward direction is more probable than any other. 

\begin{figure}[t]
\includegraphics[width=\columnwidth]{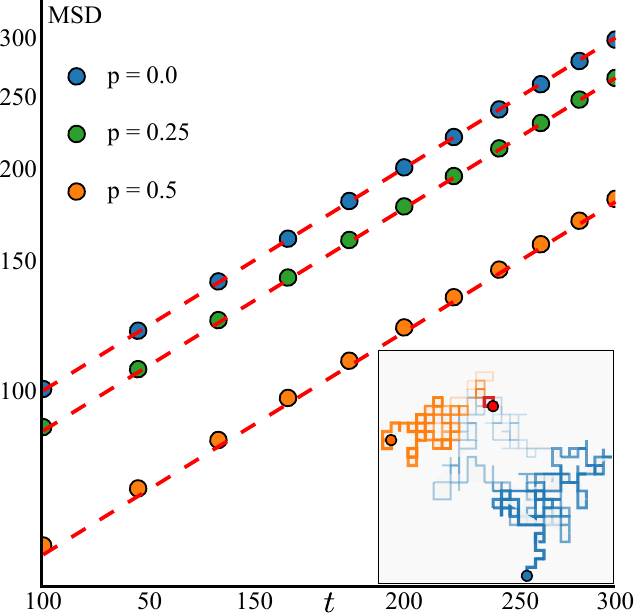}
\caption{\textbf{MSD of the chiral random walk.}
Double-logarithmic plot showing the time dependence of the mean-squared 
displacement for different values of the chirality parameter $p$. The dashed 
red lines indicate the analytical prediction of Eq.~\eqref{msd}, demonstrating 
excellent agreement with numerical simulations. The inset displays typical 
trajectories for the standard random walk (blue, $p = 0$), the CRW with 
intermediate chirality (orange, $p = 0.5$), and the deterministic chiral walk 
(red, $p = 1$). While all cases ($p<1$) exhibit Brownian scaling $\text{MSD} 
\propto t$, the bare diffusion coefficient decreases with increasing $p$ as 
given by Eq.~\eqref{diff_chiral_random_walk}. The $p=1$ limit, however, is 
deterministic and no Brownian motion is observed.}
\label{fig:msd}
\end{figure}

The mean-squared displacement (MSD) of the chiral tracer in the long-time 
limit can be straightforwardly evaluated (see appendix 
\ref{sec:app_odd_diff_tensor}) to be Brownian and is given by 
\begin{equation}
\label{msd}
    \text{MSD}(t) = D_0^p \, t,
\end{equation}
where $D_0^p$ is the bare chiral diffusion coefficient, which is given by 
\begin{equation}
\label{diff_chiral_random_walk}
    D_0^p =  \frac{a^2}{2\tau}\, \frac{1-p^2}{1+p^2}.
\end{equation}
We observe that for $p\to 1$ we have $D_0^p \to 0$, which is intuitively clear 
as the coin operator in Eq.~\eqref{coin_chiral_random} reduces to that of 
Eq.~\eqref{coin_chiral} and the motion becomes deterministic, and in circles.
In  Fig.~\ref{fig:msd}, we compare the long-time MSD of the CRW with 
Eq.~\eqref{diff_chiral_random_walk} for $p\in\{0, 0.25, 0.5\}$. Typical 
trajectories of the chiral walker and the deterministic limit of $p=1$ are 
shown in the inset.

A characteristic of the CRW is that, despite the isotropy of the system, 
chirality gives rise to tensorial diffusion. Precisely, chirality introduces 
so-called \textit{odd diffusion} \cite{hargus2021odd, kalz2022collisions, 
kalz2024oscillatory} of the walk, which is characterized by antisymmetric 
off-diagonal elements in the diffusion tensor. Relying on a Green-Kubo 
formulation for the long-time diffusion \cite{hargus2021odd} (see appendix 
\ref{sec:app_odd_diff_tensor}), the odd-diffusion tensor of the CRW can be 
obtained as
\begin{equation}
\label{odd_diffusion_tensor}
    \mathbf{D} =  D_0^p\left(\mathbf{1} + \frac{2p}{1-p^2}\, 
    \boldsymbol{\epsilon}\right),
\end{equation}
where $D_0^p$ is given in Eq.~\eqref{diff_chiral_random_walk},  $\mathbf{1} = 
\left(\begin{smallmatrix} 1 & 0 \\ 0 & 1\end{smallmatrix}\right)$ is the 
identity tensor, and $\boldsymbol{\epsilon} = \left(\begin{smallmatrix} 0 & 1 
\\ -1 & 0\end{smallmatrix}\right)$ is the antisymmetric Levi-Civita symbol 
in two dimensions. The latter represents the divergence-free elements in the 
diffusion tensor that are inherently caused by chirality and $\propto p$. Odd 
diffusion naturally appears in the overdamped, diffusive limit of physical 
systems with inherent chirality. Passive chiral systems, such as Brownian 
particles under the effect of Lorentz force \cite{chun2018emergence}, as well 
as active chiral particles \cite{muzzeddu2022active} all have an emergent 
diffusive behaviour that is characterized by a tensor of the form of 
Eq.~\eqref{odd_diffusion_tensor}, typically denoted as $\mathbf{D} = D_0^\kappa 
(\mathbf{1} + \kappa \boldsymbol{\varepsilon})$. $\kappa$ thereby is connected 
to the physical origin of chirality ($\kappa\in \mathbb{R}$, the sign denotes 
parity), e.g., for Brownian particles under the effect of Lorentz force, 
$\kappa$ is given by the ratio of Lorentz force and friction force, for active 
chiral particles it is the ratio of active chirality to rotational diffusion. 
With the mapping $p = (\sqrt{1 + \kappa^2}-1)/|\kappa|$, the CRW therefore 
models such systems on the lattice, and we note that this is only possible by 
allowing for an IDF of the walker.

\begin{figure}[t]
\includegraphics[width=\columnwidth]{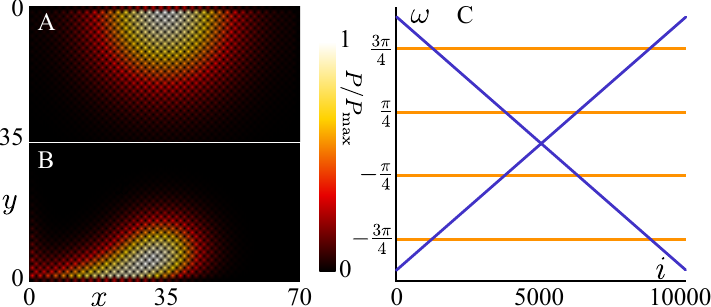}
\caption{\textbf{Emergence of topological edge modes in the chiral random walk.}
(A, B): Heatmaps of probability distribution after 300 steps on a square 
lattice with reflective boundaries, starting from a near-edge position. (A) 
Standard random walk ($p = 0$) shows symmetric diffusive spreading. (B) CRW 
($p = 0.7$) exhibits pronounced accumulation along the boundary and a 
characteristic teardrop shape, with the handedness determined by the sign of 
$p$. This edge-following behaviour, absent in the bulk, emerges from the 
antisymmetric odd-diffusion tensor and persists robustly against perturbations. 
(C) Numerically obtained quasi-energy spectrum for the CRW in the deterministic 
limit $p = 1$ with reflective boundary conditions. The bulk spectrum (orange) 
is gapped, while boundary-localized states (blue) close this gap, consistent 
with the bulk-boundary correspondence for Floquet topological insulators. 
These edge states are absent under periodic boundary conditions, confirming 
their topological origin.}
\label{fig:edge}
\end{figure}

As the characteristic antisymmetric part of the odd diffusion tensor in 
Eq.~\eqref{odd_diffusion_tensor} represents a divergence-free contribution, 
it does not affect the time-evolution of the random walk in the bulk. The 
probability distribution function (PDF) remains Gaussian, yet with a modified 
bare diffusion coefficient $D_0^p$, Eq.~\eqref{diff_chiral_random_walk}, but 
virtually indistinguishable from a nonchiral random walker. If spatial isotropy 
is broken, however, such as due to an obstacle or at a (reflecting) wall, 
chirality-induced edge-flows emerge, and already on the level of PDFs, the 
CRW can be distinguished from the ordinary random walk. In Fig. \ref{fig:edge} 
(A, B), we show a normal random walk versus a CRW close to a reflecting wall, 
where for the latter the handedness of the CRW chooses the direction of the 
characteristic teardrop shape. Within the CRW, a quantitative description of 
such edge-flow phenomena becomes possible, as the model provides a natural 
bridge between a stochastic and a unitary (quantum) evolution.

We, therefore, recall that in the limit $p=1$, the coin matrix $C_\mathrm{crw}$ 
reduces to that of the chiral coin in Eq.~\eqref{coin_chiral}, which 
specifically has the structure of a permutation matrix. That makes 
time-evolution not only bistochastic but also unitary $U^\dagger U= UU^\dagger=I$. 
In a translationally invariant systems this is reflected by an (almost) trivial 
motion: the particle is moving in circles as shown in 
Eq.~\eqref{running_in_circles}. In topologically nontrivial systems, however, 
such time evolution can be linked to an edge-flow behaviour 
\cite{hasan2010colloquium, Kitagawa2010exploring, Asboth2016short}. To illuminate 
this connection, we therefore now study the behaviour of the CRW in the $p=1$ 
limit. 

\textit{Quantum walk.} In the deterministic limit $p=1$, the evolution 
operator $U = S(I \otimes C_\mathrm{chir})$ of the CRW becomes bistochastic 
and unitary. Any unitary matrix can be interpreted as the discrete-time evolution 
operator of a quantum system \cite{nielsen2010quantum},
\begin{equation}
    \kets{\Psi(t+1)} = \mathcal{U}\kets{\Psi(t)},
\end{equation}
where in this interpretation of a two-dimensional discrete-time quantum walk, 
the state of the walker resides in a Hilbert space and can be written as
\begin{equation}
    \kets{\Psi(t)} = \sum_{\mathbf{x}, d} a_{\mathbf{x},d}(t) 
    \kets{\mathbf{x}, d}.
\end{equation}
Here $a_{\mathbf{x},d}(t)$ are the probability amplitudes associated with 
position $\mathbf{x}$ and internal state (``coin'') $d$. The unitarity of $U$ 
allows one to define an effective (Floquet) Hamiltonian via 
\begin{equation}
\label{effectiove_floquet_hamiltonian}
    H_{\text{eff}} = \frac{\mathrm{i}}{T} \log U.
\end{equation}
Such systems are periodically driven with period $T$ and are widely studied, 
especially from a topological point of view \cite{Cayssol2013floquet, 
Sajid2019creating}. A key feature of Floquet dynamics is that the quasi-energy 
spectrum of $H_{\text{eff}}$ is defined only modulo $2\pi/T$, which 
fundamentally alters the classification of topological phases. For example, it 
is found that topological invariants such as the Chern number can vanish even 
though the system is not topologically trivial \cite{Rudner2013anomalous}.

For a translationally invariant system, it is convenient to work in the 
quasi-momentum basis $\kets{\mathbf{k}} = \kets{k_x, k_y} =\sum_\mathbf{x} 
\mathrm{exp}(\mathrm{i} \mathbf{k}\cdot\mathbf{x}) \kets{\mathbf{x}}$. In this 
basis, the step operator of Eq.~\eqref{step_operator} acts diagonally on the 
internal states
\begin{subequations}
\begin{align}
    S\kets{\mathbf{k},\rightarrow} &= \mathrm{e}^{-\mathrm{i}k_x}
    \kets{\mathbf{k},\rightarrow}, \\
    S\kets{\mathbf{k},\leftarrow} &= \mathrm{e}^{\mathrm{i}k_x}
    \kets{\mathbf{k},\leftarrow},\\
    S\kets{\mathbf{k},\uparrow} &= \mathrm{e}^{-\mathrm{i}k_y}
    \kets{\mathbf{k},\uparrow},\\
    S\kets{\mathbf{k},\downarrow} &= \mathrm{e}^{\mathrm{i}k_y}
    \kets{\mathbf{k},\downarrow},
\end{align}
\end{subequations}
and the full evolution operator $U$, therefore, decomposes into $4\times 4$ blocks 
\begin{equation}
    U_\mathrm{k} = \begin{pmatrix}
        0 && 0 && \mathrm{e}^{-\mathrm{i}k_x} && 0 \\
        0 && 0 && 0 && \mathrm{e}^{\mathrm{i}k_x} \\
        0 && \mathrm{e}^{-\mathrm{i}k_y} && 0 && 0 \\
        \mathrm{e}^{\mathrm{i}k_y} && 0 && 0 && 0 
    \end{pmatrix},
\end{equation}
Given the structure of $U_\mathrm{k}$, it is convenient to consider two 
successive time steps $V_\mathbf{k}= U_\mathbf{k}^2$. The double-step operator 
$V_\mathbf{k}$ becomes block-diagonal and decomposes into two independent 
sectors characterized by the diagonal momenta $k_\nearrow = k_x + k_y$ and 
$k_\searrow = k_x-k_y$. Explicitly, 
\begin{equation}
    V_\mathbf{k} = \begin{pmatrix}
        V_{k_\nearrow} & 0 \\ 0 &V_{k_\searrow}
    \end{pmatrix},
\end{equation}
where each block takes the form 
\begin{equation}
    V_{K} = \begin{pmatrix}
        0 & \mathrm{e}^{-\mathrm{i}K} \\ \mathrm{e}^{\mathrm{i}K} & 0
    \end{pmatrix},
\end{equation}
for $K \in \{k_\nearrow, k_\searrow\}$. This decomposition maps the original 
two-dimensional quantum walk exactly onto two independent one-dimensional 
discrete-time quantum walks propagating along the lattice diagonals. 

Each one-dimensional block $V_{K}$ can be written in standard quantum-walk 
form, $V_K = \bar{S}(I\otimes\bar{C})$, with $\bar{S}$ acting as a conditional 
shift and $\bar{C}$ a unitary coin operator, which in general can be 
parametrized as
\begin{equation}
    \bar{C} = e^{-i\delta} \begin{pmatrix}\cos\theta  \ e^{i\alpha} & 
        \sin\theta  \ e^{i(\alpha+\beta)} \\ -\sin\theta  \ e^{-i(\alpha+\beta)} 
        & \cos\theta  \ e^{-i\alpha}\end{pmatrix}.
\end{equation}
We give details on the one-dimensional quantum walk in Appendix 
\ref{sec:app_dtqw}. There, we also show that the parameters of the coin 
operator $\bar{C}$ of the system at hand are $\theta = \pm\pi/2,\, \alpha = 0,\, 
\beta=\pi/2,\, \delta=\pi/2$. One-dimensional discrete-time quantum walks 
separate into two families of Hamiltonians with distinct topological phases 
for $\theta >0$ and $\theta <0$ \cite{Rudner2013anomalous}. In the present case, 
such distinct phases originate in the clockwise ($\theta = \pi/2$) or 
counterclockwise ($\theta = -\pi/2$) CRW, which, therefore, correspond to 
opposite topological phases. 

\begin{figure}
\includegraphics[width=\columnwidth]{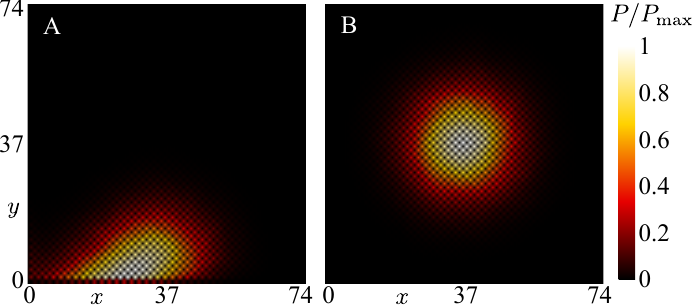}
\caption{\textbf{Robust topological edge modes in chiral random walk with 
spatially disordered chirality parameter.} Heatmaps of probability distribution 
after $300$ steps of the CRW on a square lattice, where the chirality parameter 
$p$ is randomly chosen from the interval $[0.1, 0.9]$ at each lattice site. 
Initial conditions are (A) near-edge position and (B) centre position. Despite 
the strong spatial disorder in $p$, the bulk behaviour remains diffusive and 
qualitatively unchanged, while edge currents show robustness against local 
variations in chirality strength, consistent with the topological protection 
hypothesis.}
\label{fig:randomedge}
\end{figure}

From the effective Floquet Hamiltonian of the translationally invariant 
two-dimensional CRW ($p=1$), Eq.~\eqref{effectiove_floquet_hamiltonian}, we 
find four eigenenergies for each quasi-momentum
\begin{equation}
    \omega_{k} \in \left\{-\frac{3\pi}{4},-\frac{\pi}{4},\frac{\pi}{4},
    \frac{3\pi}{4}\right\}
\end{equation}
The spectrum is gapped in the bulk but, as dictated by the bulk-boundary 
correspondence for Floquet systems, must close in the presence of boundaries 
\cite{Rudner2013anomalous}: When translational symmetry is broken by reflective 
boundaries as we are introducing them via a special coin for the boundary 
layer, Eq.~\eqref{refelctive_coin}, the quasi-energy gap closes through 
states localized at the system's edges. These states are topologically 
protected Floquet edge modes and are directly responsible for the persistent 
boundary currents observed in the deterministic chiral walk. This behaviour is 
numerically confirmed in Fig.~\ref{fig:edge}C, where we show that 
boundary-localised states close the otherwise gapped quasi-energy spectrum, as 
can be found, in contrast, with periodic boundary conditions. To present the 
robustness of these states, in Fig.~\ref{fig:randomedge} we visualize the CRW 
for a system where the $p$ value is randomly chosen at each lattice site; the 
edge states persist.

\begin{figure}[t]
\includegraphics[width=\columnwidth]{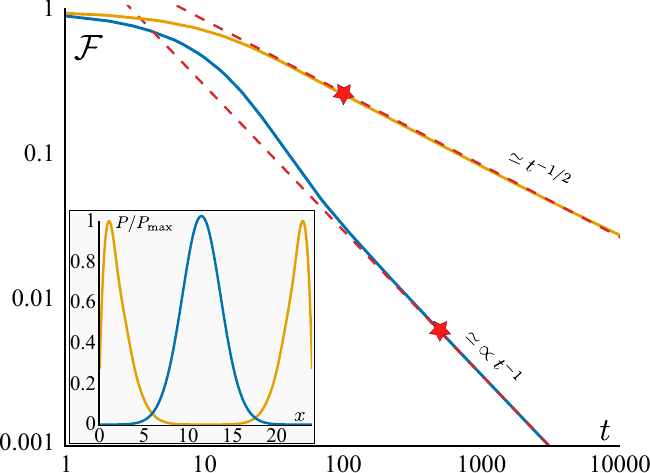}
\caption{\textbf{Edge and bulk states fidelity decay.} 
Double-logarithmic plot of the fidelity between the initial probability 
distribution and the distribution after $t$ steps of CRW evolution with 
$p = 0.9$ on a two-dimensional grid with reflective boundary conditions 
(system size $L = 500$). We compare two scenarios: an initial state taken as 
an edge eigenvector of the $p = 1$ evolution operator (orange) and a bulk 
eigenvector (blue). Edge states decay as $\simeq t^{-1/2}$, characteristic 
of effectively one-dimensional diffusion, while bulk states decay faster as 
$\simeq t^{-1}$, reflecting two-dimensional spreading. Red stars mark the 
transition from exponential to polynomial decay. Topologically protected edge 
states remain long-lived even in the dissipative regime $p < 1$. The inset 
shows the marginal probability distribution summed along one lattice direction 
(system size $L = 25$ for visibility) after 100 steps for illustration purposes 
of the initial states.}
\label{fig:fidelity}
\end{figure}

For $p<1$, the unitarity of $U$ is lost, and the evolution becomes 
non-Hermitian. Consequently, the topological edge states are no longer exact 
eigenstates of the dynamics. Nevertheless, their signatures persist as 
long-lived modes which decohere more slowly than the bulk states 
\cite{bhargava2025noise}. To view the populations of these states we compute 
the fidelity
\begin{equation}
    \mathcal{F}(t) = \bigg(\sum_\mathbf{x} \sqrt{p_\mathbf{x}(0) 
    p_\mathbf{x}(t)}\bigg)^2. 
\end{equation}
As we show in Fig.~\ref{fig:fidelity}, initial states localized on topological 
edge modes exhibit a decay $\mathcal{F}(t) \simeq t^{-1/2}$, whereas bulk 
states decay as $\mathcal{F}(t) \simeq t^{-1}$ as $t \gg 1$. A localized state 
initially decays exponentially until the system occupies all near-edge/bulk 
states, and is followed by a quasi one-dimensional random walk with decay 
$\propto t^{-1/2}$ for the edge state that effectively cuts the dimensionality 
or by a regular two-dimensional random walk with decay $\propto t^{-1}$ in the 
bulk \cite{bhargava2025noise}. We conclude that the ubiquitous phenomenon of 
robust edge modes in odd (chiral) systems \cite{souslov2019topological, 
caporusso2024phase, soni2019odd, yang2020robust, yang2021topologically, 
li2024robust, nelson2025topological} cannot only be studied in the CRW model 
but also understood from the perspective of topologically protected states in 
unitary systems. 

\begin{figure}[b]
\includegraphics[width=\columnwidth]{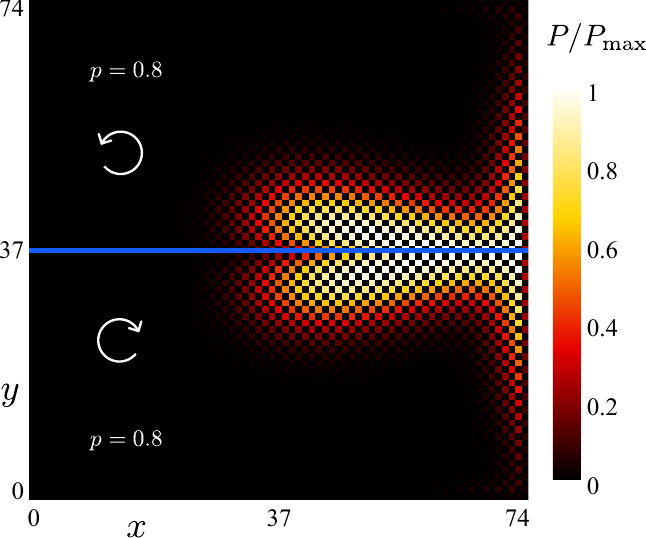}
\caption{\textbf{Topological interface currents between heterochiral domains.}
Heatmap of probability distribution after $300$ steps of the CRW in a system 
divided into two regions with opposite chirality. For $y < 37$, clockwise 
chirality with $p = 0.8$; for $y \geq 37$, counterclockwise chirality with 
$p = -0.8$ (separated by the blue line). The initial probability distribution 
was symmetrically placed at two vertices near the interface between opposite 
chirality regions. Reflective boundaries were imposed at the system edges. 
Particles accumulate along the interface and follow its path, despite the 
absence of a physical barrier, demonstrating that topological edge states form 
at the boundary between regions of opposite topological phase.}
\label{fig:leftright}
\end{figure}

\begin{figure}[t]
\includegraphics[width=\columnwidth]{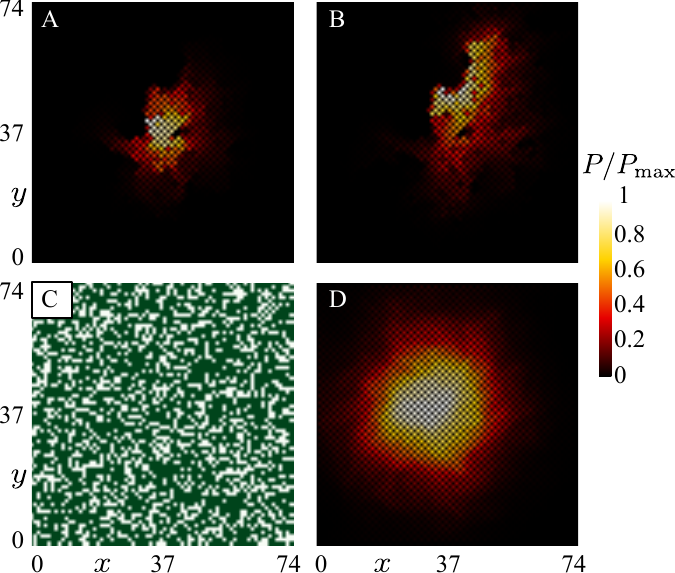}
\caption{\textbf{Enhanced transport in disordered media via chiral edge modes}
(A, B, D): Heatmaps of probability distribution after $500$ steps starting 
from the centre of a square lattice. (A) Standard random walk and (B, D) CRW 
($p = 0.8$) in the presence of disorder. In panels (A) and (B), reflective 
coins are placed at positions marked by white squares in panel (C), modelling a 
porous medium. The chiral walker (B) exhibits significantly faster spreading 
than the standard random walker (A), demonstrating enhanced transport through 
edge-following behaviour. In panel (D), the reflective obstacles are replaced 
with regions of opposite chirality (counterclockwise, $p = -0.8$) at the same 
positions. This configuration shows even more pronounced spreading, as 
interface currents along heterochiral boundaries create effective transport 
channels. (C) Random pattern of obstacle positions (white squares) used in 
panels (A), (B), and (D).}
\label{fig:porois}
\end{figure}

\textit{Applications.} In quantum walks, it is well established that 
topologically protected edge states form at interfaces between regions with 
distinct topological phases \cite{Asboth2016short}. To explore whether this 
phenomenon carries over to the CRW, we construct a system that contains two 
regions; in the upper half, we impose a counterclockwise CRW, while in the lower 
half, it is clockwise. In the deterministic limit, these two configurations 
correspond to opposite topological phases. Figure~\ref{fig:leftright} shows 
that indeed particles preferentially accumulate at the interface between the 
opposite chirality regions and follow its path, a phenomenon previously observed 
in odd systems \cite{scholz2018rotating, abdoli2020stationary}. This behaviour 
persists despite the absence of any physical barrier at the interface, 
demonstrating that the topological character of the edge states survives into 
the dissipative regime.

Having established the topological origin of edge currents in the CRW, we now 
examine whether these edge states offer practical advantages for transport. 
In quantum walks, topological edge states are known to optimize transport 
efficiency in certain geometries \cite{Asboth2016short}. To test whether also 
this phenomenon carries over to the CRW, we consider a system with uniform 
chirality ($p = 0.8$) and introduce randomly distributed reflective obstacles 
on the lattice, as shown in Fig.~\ref{fig:porois} C. These obstacles mimic a 
porous material, with reflection implemented through the reflective coin operator 
of Eq.~\eqref{refelctive_coin} at predetermined positions. The results, 
presented in Fig.~\ref{fig:porois} A and B, reveal a marked advantage of chiral 
motion: the chiral walker (Fig.~\ref{fig:porois} B) spreads significantly 
faster through the disordered medium than the standard random walker 
(Fig.~\ref{fig:porois} A). This enhanced transport can be attributed to the 
edge-following behaviour, which allows the chiral tracer to navigate around 
obstacles more efficiently. Remarkably, we find that transport is enhanced even 
further when the reflective obstacles are replaced by regions of opposite 
chirality (Fig.~\ref{fig:porois} D). In this configuration, the interface 
currents discussed in Fig. \ref{fig:leftright} appear to guide the particle 
along the boundaries of the heterochiral regions, creating effective transport 
channels through the disordered landscape. While these examples represent 
preliminary explorations, they demonstrate the predictive power of the CRW 
model and suggest promising avenues for discovering new transport phenomena 
in odd systems. The ability to tune chirality locally opens possibilities for 
engineering transport pathways in soft matter and colloidal systems without 
relying on external fields or confinement.

\textit{Conclusions.} We have established the CRW as a minimal microscopic 
model that captures the essential phenomenology of odd diffusion while 
providing a direct bridge to topological quantum walks. By introducing an IDF 
to the classical random walk and a single tuneable parameter $p$, we interpolate 
continuously between standard diffusive motion ($p = 0$) and deterministic, 
topologically nontrivial quantum walks ($p = 1$).

The central finding of this work is that the topological protection 
characteristic of the unitary limit persists remarkably into the dissipative 
regime. Edge currents in chiral systems, previously understood primarily 
through continuum hydrodynamics, emerge naturally in our discrete framework as 
manifestations of Floquet topological edge states. This connection is more 
than a formal analogy: the bulk-boundary correspondence of quantum systems 
provides quantitative predictions for the robustness and decay properties of 
edge modes for classical odd-diffusive tracers. Even in the presence of 
dissipation ($p < 1$) and spatial disorder, edge states decay as $\simeq 
t^{-1/2}$, significantly slower than bulk modes, which follow $\simeq t^{-1}$, 
directly reflecting the spectral structure of the underlying effective 
Hamiltonian.

Beyond theoretical unification, the CRW offers practical predictive power 
for complex transport scenarios. Our simulations demonstrate enhanced diffusion 
through porous media and the emergence of interface currents between 
heterochiral domains---phenomena that would be challenging to anticipate from 
hydrodynamic equations alone. The local tunability of chirality in the model 
suggests new design principles for transport control in soft matter systems. 
For instance, patterned regions of opposite chirality could serve as transport 
channels, guiding particles along predetermined paths without physical 
confinement or external forcing.

The correspondence we have established allows the extensive toolkit of 
topological band theory to be harnessed for classical dissipative systems. 
Questions about transport in complex geometries, the role of quenched disorder, 
or the effects of heterogeneous chirality patterns can now be addressed 
through well-established quantum mechanical methods. This opens pathways for 
the rational design of chiral colloidal devices and active matter systems with 
engineered transport properties.

Looking forward, the CRW model invites experimental validation in systems 
where local chirality can be controlled, such as colloidal particles in rotating 
magnetic fields or active matter with tunable chirality \cite{mecke2024emergent}. 
Extensions to other lattice geometries \cite{mallory2019activity, wang2024robo}, 
or the incorporation of self-propulsion \cite{shaebani2020computational}, 
interactions between multiple walkers \cite{mason2023exact}, or obstacles and 
disorder \cite{reinken2020organizing, chan2024chiral, mecke2025obstacle} 
represent natural next steps. The framework may also prove valuable in 
understanding topological phenomena in other non-Hermitian classical systems 
\cite{ashida2020non, sone2026hermitian}, where the interplay between topology 
and dissipation remains an active frontier.

\textit{Acknowledgments.} J. W. acknowledges support by the Narodowe Centrum 
Nauki (OPUS grant No. 2024/53/B/ST2/04103). We acknowledge fruitful discussions 
with Abhinav Sharma, Ralf Metzler and Antoni W{\'o}jcik.

\section*{Data Availability Statement}
The data that support the findings of this article are openly available at 
\cite{data}

\appendix

\section{Odd diffusion tensor}
\label{sec:app_odd_diff_tensor}

The (chiral) random walk follows a discrete-time evolution with timestep 
$\tau$ on a two-dimensional square lattice $a\mathbb{Z}^2$, where $a$ is 
the lattice constant. The internal (coin) space $\mathcal{D}$ has $z$ states 
labelled by $d=1, \ldots, z$. 
 
Each coin state is associated with a unit displacement vector $\mathbf{e}_d 
\in \mathbb{R}^2$ that points to the lattice neighbour reached when the walker 
is in coin state $d$. The time-evolution $U$ of the random walk can be 
decomposed as $U=S(I \otimes C)$ into a coin operator (a classical stochastic 
matrix) $C \in \mathbb{R}^{z\times z}$, and a conditional step operator $S$ 
which moves the walker by $a \mathbf{e}_d$. The coin chain is a Markov Chain 
with transition matrix $C$. Assume $C$ is irreducible and aperiodic, so it has 
a unique stationary distribution $\boldsymbol{\pi}$ satisfying 
$C\boldsymbol{\pi} = \boldsymbol{\pi}$. Following Ref.~\cite{gilbert2009diffusion}, 
we can write the diffusion tensor via a discrete analogue of the 
Green-Kubo formula as 
\begin{equation}
\label{discrete_diffusion_tensor}
    \mathbf{D} = \frac{a^2}{2\tau} \frac{1}{2} \left[\mathbf{C}_v(0) 
    + 2\sum_{t=1}^\infty \mathbf{C}_v(t)\right],
\end{equation}
where $\mathbf{C}_v(t) = \langle (\mathbf{v}_0 - \overline{\mathbf{v}}) 
\otimes (\mathbf{v}_n - \overline{\mathbf{v}})^\mathrm{T}\rangle$ is the 
velocity autocorrelation tensor and $\mathbf{v}_t$ is the walker's ``velocity'' 
at step $t$, $\mathbf{v}_t = \mathbf{e}_{d_t}$, where $d_t$ is the coin state 
at time $n$. Further, $\overline{\mathbf{v}}= \sum_d \pi_d \mathbf{e}_d$ is the 
stationary mean velocity, i.e., $\overline{\mathbf{v}} = \mathbf{0}$ in the 
unbiased case. $\langle \cdot \rangle$ denotes expectation in the stationary 
measure. Note that $\mathbf{C}_v(0) = \mathbf{1}$. 

Because the coin chain is Markov, the joint distribution of $(d_0, d_t)$ at 
stationarity is $\mathrm{Prob}\{d_0 = i, d_t =j\} = \pi_i [C^t]_{ij}$. The 
velocity autocorrelation tensor in the unbiased case is thus given by
\begin{equation}
\mathbf{C}_v(t) = \sum_{i, j=1}^z \pi_i [C^t]_{ij}\, (\mathbf{e}_i - 
\overline{\mathbf{v}}) \otimes (\mathbf{e}_j - \overline{\mathbf{v}})^\mathrm{T}.
\end{equation}
From here, we can write a numerically convenient form for calculating the 
diffusion tensor. Define the ``lattice-geometry'' matrix $\mathbf{E} \in 
\mathbb{R}^{2\times z}$, whose $d$th column is $\mathbf{e}_d$. Put the 
stationary distribution elements on the diagonal as $\boldsymbol{\Pi} = 
\mathrm{diag}(\boldsymbol{\pi})$ and define the centred geometry $\mathbf{V} = 
\mathbf{E} - \overline{\mathbf{v}}\,  \mathbf{1}_z^\mathrm{T}$, where 
$\mathbf{1}_z$ is the $z$-vector of ones. The $d$th row of $\mathbf{V}$ thus 
represents the centred velocity for the $d$th coin degree, i.e., 
$\mathbf{V}_{:, d} = \mathbf{e}_d - \overline{\mathbf{v}}$. The diffusion 
tensor of Eq.~\eqref{discrete_diffusion_tensor} thus can be brought into the 
explicit $2 \times 2$ form 
\begin{equation}
\label{discrete_diffusion_tensor_2}
    \mathbf{D} = \frac{a^2}{2\tau} \frac{1}{2} \left[\mathbf{1} 
    + 2\sum_{t=1}^\infty \mathbf{V}\boldsymbol{\Pi} [C^t]^\mathrm{T} 
    \mathbf{V}^\mathrm{T}\right].
\end{equation}
Given the basic random walk ingredients of $(i)$, the coin operator $C$, from 
which a stationary vector $\boldsymbol{\pi}$ can be computed and $(ii)$, the 
list of lattice geometry vectors $\{\mathbf{e}_d\}$ (i.e., the conditional 
step operator $S$), Eq.~\eqref{discrete_diffusion_tensor_2} thus gives access 
to the long-time diffusive properties of the walk. 

We can specify Eq.~\eqref{discrete_diffusion_tensor_2} to the CRW on the 
square lattice with coin operator given by Eq.~\eqref{coin_chiral_random} of the 
main text. Chirality contributes via a permutation matrix and therefore does 
not alter the stationary distribution, which is given by $\boldsymbol{\pi} = 
(1/4,1/4,1/4,1/4)^\mathrm{T}$. Thus, $\boldsymbol{\pi}$ is uniform and 
independent of $p$. The square lattice is characterized by the direction 
vectors $\mathbf{e}_1=(1, 0)$, $\mathbf{e}_2 = (-1, 0)$, $\mathbf{e}_3=(0,1)$ 
and $\mathbf{e}_4=(0,-1)$ according to the ordering of our coin degrees 
$\mathcal{D}=\{\rightarrow, \leftarrow, \uparrow, \downarrow\}$ and the step 
operator in Eqs.~\eqref{step_operator_1}-\eqref{step_operator_4}. One obtains 
straightforwardly
\begin{subequations}
\begin{align}
    C_v^{xx}(t) &= C_v^{yy}(t) = 2\cos\left(t\, \pi/2\right)\ p^t,\\
    C_v^{xy}(t) &= 2\cos\left((t-1)\, \pi/2\right)\ p^t,\\
    C_v^{yx}(t) &= 2\cos\left((t+1)\, \pi/2\right)\ p^t,
\end{align}
\end{subequations}
from which it follows that 
\begin{subequations}
\begin{align}
    D_{xx} = D_{yy} &= \frac{a^2}{2\tau}\, \frac{1-p^2}{1+p^2},\\
    D_{xy} = - D_{yx} &= \frac{a^2}{2\tau}\, \frac{2p}{1+p^2}.
\end{align}
\end{subequations}

\section{One dimensional discrete time quantum walk}
\label{sec:app_dtqw}
The discrete-time quantum walk describes the dynamics of a particle on a 
given lattice equipped with an internal degree of freedom, for a review see 
Ref.~ \cite{Kempe2003quantum}. For a one-dimensional walk, we consider a 
particle with position $x \in \mathbb{Z}$ and a two-dimensional internal 
state space, the so-called coin degree of freedom. The particle's wave function 
can be written as
\begin{equation}
    \kets{\psi} = \sum_x \kets{x} \otimes \left(a_x\kets{+1} + 
    b_x\kets{-1}\right),
\end{equation}
where $a_x$ and $b_x$ are the probability amplitudes, $\kets{x}$ denotes 
the position basis state, and $\kets{\pm1}$ represent the coin basis states.

The time evolution of the discrete-time quantum walk is governed by a 
unitary evolution operator $\kets{\psi(t + 1)} = U\kets{\psi(t)}$, which 
we decompose as
\begin{equation}
    U = S \left( \sum_x |x\rangle\langle x| \otimes C_x\right),
\end{equation}
where $S$ is the step operator and $C_x$ is the (in general position-dependent) 
coin operator. The step operator acts as a state-dependent translation given by
\begin{equation}
    S\kets{x,\pm1} = \kets{x\pm1,\pm1},
\end{equation}
while the coin operator performs a unitary transformation on the internal 
degree of freedom. The most general form of the coin operator can be 
parametrized as
\begin{equation}
    C_x = e^{-i\delta_x} \begin{pmatrix}e^{i \zeta_x}\cos\theta_x 
        & e^{i (\zeta_x+\sigma_x)}\sin\theta_x\\-e^{-i (\zeta_x+\sigma_x)}
        \sin\theta_x & e^{-i \zeta_x}\cos\theta_x\end{pmatrix},
\end{equation}
where with real parameters $\delta_x,\, \zeta_x,\, \sigma_x, \, \theta_x$.

To make contact with solid-state theory, we introduce an effective Hamiltonian 
via $H_{\text{eff}} = \mathrm{i}\log U/T$, with eigenenergies defined through 
$U\kets{\psi} =\mathrm{exp}(-\mathrm{i}\omega)\kets{\psi}$. For translationally 
invariant systems, where the coin operator is position-independent, we work in 
the quasi-momentum basis $\kets{k} = \sum_x \mathrm{exp}(-\mathrm{i}kx)
\kets{x}$ with $k$ restricted to the first Brillouin zone $-\pi \leq k \leq 
\pi$. In this basis, the step operator acts diagonally as
\begin{equation}
    S\kets{k,\pm1} = \mathrm{e}^{\mp \mathrm{i}k}\kets{k,\pm1},
\end{equation}
while the coin operator remains unchanged since it acts only on the internal 
degree of freedom. We now treat $\theta$ as a free parameter and define a 
family of effective Hamiltonians
\begin{equation}
    H(\theta) = \oint_{BZ}\mathrm{d}k \left(\kets{k}\bras{k}\otimes \bar{H}_k \right),
\end{equation}
where $\bar{H}_k$ is the $2 \times 2$ Bloch Hamiltonian. This can be 
represented as $\bar{H}_k = \delta I + \omega_k \mathbf{n}_k \cdot 
\boldsymbol{\sigma}$, with $\boldsymbol{\sigma} = (\sigma_x, \sigma_y, \sigma_z)$ 
denoting the vector of Pauli matrices and the normalized Bloch vector given by
\begin{equation}
    \mathbf{n}_k = \frac{1}{\sin\omega_k}\begin{pmatrix}\sin\theta\sin(k-\tau) 
        \\ -\sin\theta\cos (k-\tau)\\ \cos\theta\sin (k-\alpha)\end{pmatrix},
\end{equation}
where $\tau = \alpha + \beta$. The quasi-energy $\omega_k$ satisfies
\begin{equation}
    \cos\omega_k = \cos\theta\cos(k-\alpha),
\end{equation}
and the eigenenergies are given by $\omega_{k\pm} = \delta \pm \omega_k$, 
with corresponding projectors
\begin{equation}
    \rho_{k\pm}=\frac{1}{2}(I + \mathbf{n}_\pm \cdot\boldsymbol{\sigma}),
\end{equation}
and $\mathbf{n}_{k\pm} = \pm\mathbf{n}_k$.

For the CRW in the deterministic limit $p = 1$, we substitute the coin 
parameters $\alpha = 0$, $\beta = \pi/2$, and $\delta = \pi/2$ (with $\theta$ 
remaining a free parameter). This yields the normalized Bloch vector

\begin{equation}
\mathbf{n}_k = \frac{1}{\sin\omega_k} \begin{pmatrix}-\sin\theta\cos k 
\\ -\sin\theta\sin k\\ \cos\theta\sin k\end{pmatrix},
\end{equation}
and the quasi-energy satisfies
 \begin{equation}
     \omega_k =\arccos\big(\cos\theta\cos k\big),
 \end{equation}
The eigenenergies of the effective Hamiltonian thus take the form 
$\omega_{k\pm} = \pi/2 \pm \omega_k$. 

From this expression, we observe that the spectrum of the effective 
Hamiltonian can be gapped. Let $\mathcal{S}$ denote the subset of $\theta$ 
for which the Hamiltonian exhibits a spectral gap. This set is a disconnected 
topological space with partition $\mathcal{S} = \mathcal{S}_{\theta < 0} 
\cup \mathcal{S}_{\theta > 0}$, where $\mathcal{S}_{\theta < 0} = \{\theta 
\mid -\pi < \theta < 0\}$ and $\mathcal{S}_{\theta > 0} = \{\theta \mid 0 < 
\theta < \pi\}$. Grudka et \textit{al.}~\cite{Grudka2023topological} 
rigorously proved that $\mathcal{S}_{\theta > 0}$ and $\mathcal{S}_{\theta < 0}$ 
support distinct topological phases, using the notion of homotopic relative 
maps from the Brillouin zone to the Bloch sphere. For our CRW, the clockwise 
($\theta = \pi/2$) and counterclockwise ($\theta = -\pi/2$) configurations 
therefore correspond to opposite topological phases, which underlie the edge 
state formation at interfaces between heterochiral domains discussed in the 
main text.

\end{document}